\documentclass[conference]{IEEEtran}
\IEEEoverridecommandlockouts
\usepackage{cite}
\usepackage{amsmath,amssymb,amsfonts}
\usepackage{algorithmic}
\usepackage{graphicx}
\usepackage{textcomp}
\usepackage{xcolor}
\usepackage[caption=false,font=footnotesize]{subfig}
\usepackage{multirow}
\usepackage{url}
\usepackage{hyperref}
\usepackage{flushend}
\usepackage{fancyhdr}
\makeatletter
\let\ps@headings\ps@fancy
\let\ps@IEEEtitlepagestyle\ps@fancy
\makeatother

\def\BibTeX{{\rm B\kern-.05em{\sc i\kern-.025em b}\kern-.08em
    T\kern-.1667em\lower.7ex\hbox{E}\kern-.125emX}}

\begin{document}

\title{On the Vulnerabilities and Data-Free Removal of Watermarks in Batch Normalization Layers
\thanks{Work partially funded by the Spanish Ministry of Science, Innovation and Universities via a doctoral grant to the first author (FPU22/01929), and by Xunta de Galicia and the European Regional Development Fund, under project ED431C 2025/41.}
}
\author{\IEEEauthorblockN{Elena Rodríguez-Lois, Fernando Pérez-González}
\IEEEauthorblockA{
\textit{Signal Theory and Communications Department}\\
atlanTTic Research Center, University of Vigo, E.E. de Telecomunicación, 36310 Vigo, Spain\\
{erodriguez, fperez}@gts.uvigo.es\\\vspace{-8mm}}
}

\maketitle

\begin{abstract}
Batch Normalization (BN) layers have become a common choice for embedding white-box watermarks in Federated Learning (FL), providing a simple and effective mechanism for model ownership verification and participant identification. However, previous works have focused on robustness against generic model modifications, leaving the security implications of this design largely unexplored. In this paper, we identify fundamental vulnerabilities of BN-based watermarking schemes and propose a practical data-free attack that exploits them. Experimental results on two representative FL watermarking schemes show that the proposed attack consistently removes the watermark with minimal impact on model performance, even without access to representative training data. More broadly, our findings highlight the importance of evaluating DNN watermarking schemes under threat models consistent with Kerckhoffs's principle, rather than relying on security through obscurity.
\end{abstract}

\begin{IEEEkeywords}
DNN Watermarking, Federated Learning, White-Box, Batch Normalization, Attack, Data-Free
\end{IEEEkeywords}

\section{Introduction}

Deep learning applications have become increasingly popular in recent years, achieving unprecedented performance across a wide range of tasks. With this, the high cost of data collection and model training has made them valuable intellectual property, vulnerable to theft and unauthorized use. In the context of Federated Learning (FL), where multiple data-owners collaborate to train a model without sharing their private data, watermarking has become an essential tool to establish model ownership~\cite{Tekgul21,Li26}, verify the ownership rights of individual data-owners~\cite{Zhang24,Bowen_Li22,WYang23,Chao24}, and even detect specific malicious participants through the use of unique individual watermarks~\cite{Shao23,Chen25,Yang26,Karim26}. Depending on the nature of the verification, a watermark can be white-box, if one needs to inspect the model parameters to verify its presence, or black-box, if it is encoded in input-output behavior and can be detected through queries to the model. 

Among existing approaches, embedding white-box watermarks into the scaling parameters of Batch Normalization (BN) layers has become common practice in FL~\cite{Bowen_Li22,Chao24,Shao23,WYang23,Chen25,Yang26,Karim26}. Previous works have highlighted how this design allows for a simple embedding, with limited impact on main task accuracy and improved robustness to fine-tuning attacks~\cite{Bowen_Li22}, given that batch statistics remain mostly unchanged. It has also been a straightforward way to separate black and white-box watermarks~\cite{Shao23}, as the former typically avoid modifying BN layers to preserve main-task performance. Unfortunately, this choice also presents some vulnerabilities that, to the best of our knowledge, have not yet been discussed in the literature, raising important concerns about the security of BN-based watermarking schemes.

\begin{figure}[t]
    \centering
    \includegraphics[width=\columnwidth]{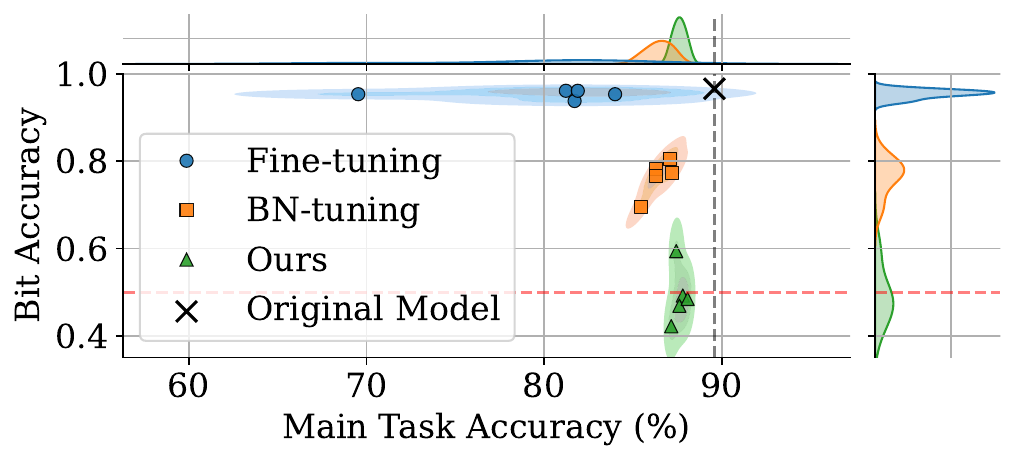}
    \caption{Comparison of potential removal attacks in terms of main task accuracy and watermark bit accuracy for VGG16 trained on CIFAR-10 with FedTracker~\cite{Shao23}. Dashed lines represent the random guessing of the watermark (red) and average original performance of the model before the attack (gray).}
    \label{fig:attack_comparison}
\end{figure}

In this paper, we investigate these vulnerabilities and demonstrate that they can be exploited to perform an effective and data-free watermark removal attack. Our main contributions are as follows:
\begin{enumerate}
    \item To identify and analyze previously overlooked vulnerabilities of BN-based watermarks.
    \item To propose a data-free watermark removal attack that exploits these vulnerabilities, enabling the theft of a watermarked classification model while preserving its main-task performance, even without access to representative training data.
\end{enumerate}

The rest of the paper is organized as follows: The vulnerabilities of BN watermarks are first discussed in Section~\ref{sec:vulnerabilities}. Then, Section~\ref{sec:proposed} presents our proposed data-free removal attack, with the experimental results validating its effectiveness in Section~\ref{sec:experimental}. Finally, Section~\ref{sec:conclusions} concludes the paper and discusses the main findings and recommendations. Our implementation is available for reproducibility at \url{https://github.com/erodriguezlois/BN-DF-Attack}.

\section{Vulnerabilities of BN Watermarking}\label{sec:vulnerabilities}

BN layers are a key component in DNNs, as they optimize the distribution of hidden feature channels, improving convergence. In each BN layer $l$, input activations $\mathbf{A}_l^{B\times C\times H\times W}$ are normalized using the running batch mean $\hat{\mu}_{l,c}$ and standard deviation $\hat{\sigma}_{l,c}$ of each feature channel $c$, and then adjusted according to parameters $\gamma_{l,c}$ and $\beta_{l,c}$ to obtain $\mathbf{A}'_l$, as
\begin{equation}\label{eq:bn}
A'_{l, b, c, h, w} = \gamma_{l,c} \left( \frac{A_{l,b, c, h, w} - \hat{\mu}_{l,c}}{\sqrt{\hat{\sigma}_{l,c}^2 + \epsilon}} \right) + \beta_{l,c}.
\end{equation} 
BN-based watermarking typically embeds secret bits into the scaling parameters $\boldsymbol{\gamma}_l$ via an additional regularization term that ensures accurate extraction.

While effective, this approach also raises critical security concerns. As part of an early discussion of the main challenges and pitfalls facing DNN watermarking, Barni \emph{et al.}~\cite{Barni21} emphasized the importance of adopting an appropriate security model against adversaries attempting to remove the watermark. Nevertheless, many current schemes still overlook this aspect, assuming attackers may not even be aware of the presence of the watermark nor the mechanisms involved in the embedding and verification. Consequently, robustness is often evaluated only against blind attacks, ignoring Kerckhoffs's principle, where a security system should remain secure assuming all aspects of the design are public knowledge except for a secret key~\cite{Kerckhoffs}. This approach to security has been criticized since early digital watermarking literature~\cite{Barni03}, but it becomes more problematic in FL, where potential attackers have likely agreed upon the defensive measures themselves, and may even participate in the embedding of their own watermarks~\cite{Bowen_Li22}.

\subsection{The Problem of Detectability}\label{subsec:detectability}

\begin{figure}[ht]
    \centering
    \subfloat[VGG16 with FedTracker]{\includegraphics[width=0.5\columnwidth]{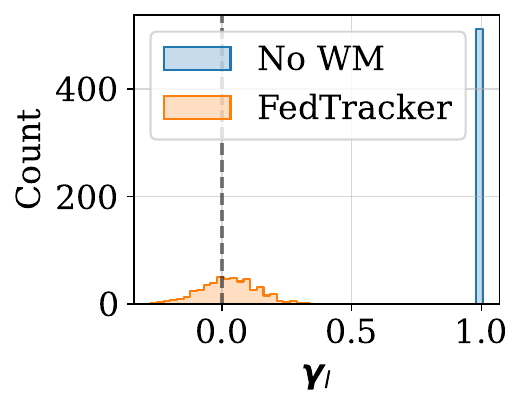}}
    \subfloat[AlexNet with FedIPR]{\includegraphics[width=0.5\columnwidth]{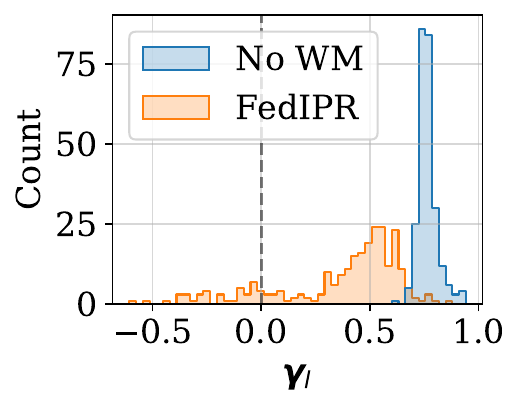}}\\
    \caption{Comparison of scaling parameter $\boldsymbol{\gamma}_l$ distributions of a specific BN layer for clean and watermarked models.}
    \label{fig:detectability}
\end{figure}

While initial works on DNN watermarking still focused on the problem of detectability~\cite{Cortinas-Lorenzo20}, many recent works have neglected this requirement. In the case of BN watermarks, and aside from specific indicators such as Passport layers~\cite{Bowen_Li22}, their presence is easily detectable. Figure~\ref{fig:detectability} illustrates this behavior for models trained with two representative BN watermarking schemes: FedTracker~\cite{Shao23} and FedIPR~\cite{Bowen_Li22}. The distributions correspond to the scaling parameters in the last BN layer of VGG16 and the BN in the final Passport layer of AlexNet, respectively. In non watermarked networks, scaling parameters of BN layers tend to remain close to their original initialization~\cite{Davis22} and stay positive~\cite{Liu17}: A sign flip would effectively cancel the gradients flowing through the corresponding feature channel as the parameter crosses zero, potentially preventing it from receiving further updates. In contrast, optimizing for an additional objective, such as white-box watermarking, can incentivize sign changes in the scaling parameters. In such case, negative values would be perfectly feasible, as a zero crossing would only be a temporary state during training. The potential detection makes it more unlikely for an adversary to be unaware of the presence and nature of the watermark.

\subsection{The Problem of Targeted Attacks}\label{subsec:targeted_attacks}

If obscurity is not an option to defend BN watermarks, an adversary would not necessarily have an incentive to fine-tune the full model, which has been the baseline attack (alongside parameter pruning) in the previously mentioned works~\cite{Bowen_Li22,Chen25,Shao23,WYang23,Yang26,Karim26} and shown to be unproductive. Notably, a naive BN fine-tuning (BN-tuning) with the data-owner's own data would be a more effective attack, as shown by Figure~\ref{fig:attack_comparison}.\footnote{In Figure~\ref{fig:attack_comparison}, the fine-tuning attack performs 50 epochs of the data-owner's local dataset with the original training settings, the BN-tuning changes only the BN parameters, with a greater learning rate by a factor of 100, and the proposed scheme was performed as described in Section~\ref{sec:proposed} and~\ref{sec:experimental}.} Because BN-tuning modifies only a small subset of the network parameters, it is less prone to overfitting on the limited local dataset, allowing the use of a greater learning rate to disturb the original parameters without considerably compromising the performance on the main task. Building on this intuition, the attack proposed in this work more directly targets the BN scheme, achieving complete watermark removal.

\subsection{The Problem of Batch Normalization Statistics}

Batch normalization layers store statistics that would allow any malicious participant, even without access to main task data, to reproduce the distributions expected at their input. With this advantage, one can easily isolate each batch normalization layer from the rest of the network and compute a useful re-initialization as a linear problem matching the expected mean and variation, regardless of the nonlinearities in the overall main task function. While it is likely that this re-initialization itself is not an optimal operating point for the main task, the preservation of approximate statistics is still necessary to properly reuse the remaining trained layers of the network (see Section~\ref{subsec:exp_initialization}). 
Considering this, schemes based on custom normalization layers such as~\cite{Fei23} would not be as easily attacked by a data-free adversary.

\subsubsection{Exact or Approximate Statistics}
\begin{figure}[ht]
    \centering
    \includegraphics[width=\columnwidth]{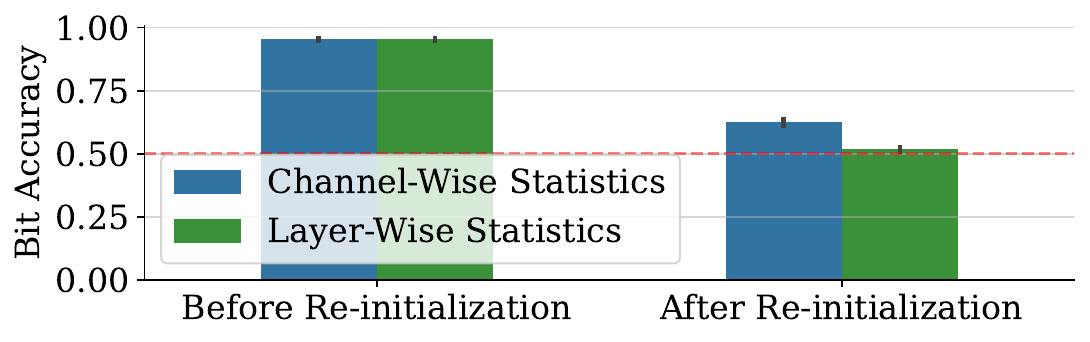}
    \caption{Comparison of potential re-initialization approaches for the BN parameters, in terms of watermark bit accuracy, for the models trained with FedIPR in Section~\ref{sec:experimental}. The red dashed line represents random guessing. }
    \label{fig:approximate_vs_exact}
\end{figure}

Although it may be tempting to replicate the exact output mean and variance for each feature channel (e.g., setting the scale parameter to the absolute value), this also poses its own risks: In works where the BN watermark identifies the global model~\cite{Bowen_Li22}, the joint training of the main task alongside the watermark will inadvertently spread the effects of the embedding beyond the BN layers, meaning that replicating the exact distributions may also recover part of the watermark information. This effect can be observed in Figure~\ref{fig:approximate_vs_exact}, where replicating channel-wise statistics leads to a watermark bit accuracy well above random guessing. Additionally, even for FL traitor tracing schemes with individual watermarks for different model copies, where the collaborative nature of FL implies the remaining layers are able to solve the task for any of the data-owners' BN watermark, a knowledgeable defender might be able to reverse such an attack, multiplying by a sign vector of the protected model.\footnote{This was validated experimentally with FedTracker~\cite{Shao23}: even when the sign vector was extracted from a different model copy (from a non-guilty participant) a defender was still able to recover the correct watermark.} On the other hand, a layer-wise re-initialization such as the one described in Section~\ref{subsec:reinitialization} would be more desirable to remove the watermark.

\section{A Data-Free Attack on BN Watermarks}\label{sec:proposed}

\begin{figure}[ht]
    \centering
    \includegraphics[width=\columnwidth]{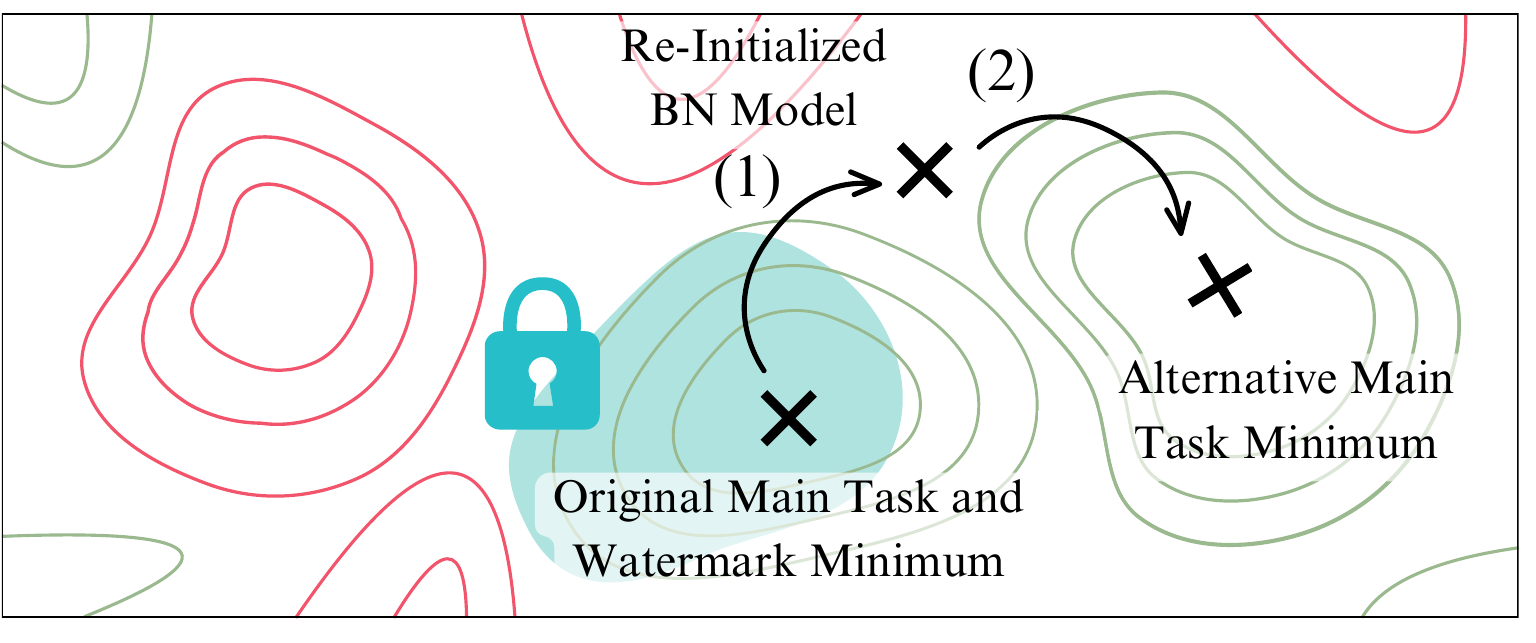}
    \caption{Symbolic diagram depicting the removal attack: The main task loss landscape (contour) has a large number of local minima (green) independent of the initial watermarking minimum (blue). From the original model parameters one can re-initialize the BN layers to lose the watermark information (1) and BN-tune the classification task to mitigate the impact on the model's performance (2).}
    \label{fig:removal_diagram}
\end{figure}

Considering the vulnerabilities described in Section~\ref{sec:vulnerabilities}, this Section presents a data-free attack able to remove BN watermarks with limited impact on the main task performance. Because of its data-free nature, this attack is suitable for a wide range of malicious attackers, including \textit{freeloaders} (participants contributing little or no useful data during training), aggregators (central servers coordinating training and update aggregation), and data-owners with non IID distributions (see Section~\ref{subsec:exp_noniid}). In a nutshell, the attack is performed in two steps, as depicted in Figure~\ref{fig:removal_diagram}:
\begin{enumerate}
    \item \textbf{Replicating Approximate BN Statistics}: Replacing the BN parameters with a suitable re-initialization, obtaining a model close to the original, but independent of the BN watermark.
    \item \textbf{Finding an Alternative Local Minimum for Classification}: Optimize the BN parameters for the main task using proxy samples extracted from the original network, independently of the original watermarking objective. 

\end{enumerate}

\subsection{Replicating Approximate BN Statistics}\label{subsec:reinitialization}

Considering~\eqref{eq:bn}, one can randomly generate artificial samples that resemble the expected distribution at the input as ${A_{l, \text{proxy}, b, c, h, w} \sim \mathcal{N}\left(\hat{\mu}_{l, c}, \, \hat{\sigma}_{l, c}^2\right)}$, and use the resulting $\mathbf{A}'_{l,\text{proxy}}$ from~\eqref{eq:bn} to estimate the expected output distribution. Instead of replicating the statistics for each channel $c$, the layer-wise mean and standard deviation of $\mathbf{A}'_{l,\text{proxy}}$ can be used to re-initialize $\boldsymbol{\gamma}_l$ and $\boldsymbol{\beta}_l$ as
\begin{equation}
    \boldsymbol{\gamma}_l = \operatorname{std}_{b,c,h,w}(\mathbf{A}'_{l,\text{proxy}}) \cdot \mathbf{1}_{C} \quad \boldsymbol{\beta}_l = \operatorname{mean}_{b,c,h,w}(\mathbf{A}'_{l,\text{proxy}}) \cdot \mathbf{1}_{C},
\end{equation}
where $\mathbf{1}_{C} \in \mathbb{R}^{C}$ represents a column vector of ones matching the number of channels $C$ in each BN layer. 

\subsection{Finding an Alternative Local Minimum for Classification}

The re-initialed model is likely not an optimal solution for the main task: not only is the relationship between feature channels disturbed, but sign changes in particular can cause a significant impact due to ReLU activations. Because of this, the new BN parameters need to be adjusted to take best advantage of the trained layers. While this second step does not explicitly guide the model away from the original watermark, which would require knowledge of the watermark itself, the combination of re-initialization and DNN over-parametrization makes it extremely unlikely that the model will return to its original minimum. The data-free options for this step are limited, as loss functions emulating the behavior of the original model may inadvertently also recover some watermark information.\footnote{This was seen experimentally when comparing BN-tuning with an unlabeled non IID dataset through Kullback-Leibler divergence loss.} Remarkably, it is possible to extract representative data points for the different classes $\mathcal{C}$ in the main task from the trained network, and use their corresponding labels to BN-tune the re-initialized model into a local minimum near the original.

Following the intuition in previous works~\cite{Yin20} that exploited BN statistics to recreate training data, an attacker could start from a batch of $B$ samples $\mathbf{X}^{
B\times C\times H\times W}$ with IID draws from $\mathcal{U}(0,1)$, together with a random label vector $\mathbf{y} \in \mathcal{C}^B$, and optimize this noise on the original model to act as a substitute dataset.\footnote{There are many possible solutions for this step, many of which may be more effective. In this work we consider a simple approach to showcase the effect of the data-free attack itself, but any good distillation function could potentially be used to improve the performance of the stolen model.} We assume that a data-free attacker would also have access to normalization statistics $\boldsymbol{\mu}_{\text{D}}$ and $\boldsymbol{\sigma}_{\text{D}}$ necessary to normalize any input to the model, and so a first step is to optimize this noise distribution through
\begin{equation}
    \mathcal{L}_{\text{D}}(\mathbf{X}) = \| \boldsymbol{\mu}_{\mathbf{X}} - \boldsymbol{\mu}_{\text{D}} \|_2 + \|  \boldsymbol{\sigma}_{\mathbf{X}} - \boldsymbol{\sigma}_{\text{D}}\|_2.
\end{equation}
Considering now for each BN layer of the model $l$, the current batch statistics for a given input $\boldsymbol{\mu}_l(\cdot)$ and $\boldsymbol{\sigma}_l(\cdot)$, and the previously learned running statistics $\hat{\boldsymbol{\mu}}_l$ and $\hat{\boldsymbol{\sigma}}_l$, an attacker can also optimize 
\begin{equation}
    \mathcal{L}_{\text{BN}}(\overline{\mathbf{X}}) = \sum_{l \in \mathcal{L}} \left( \|\boldsymbol{\mu}_l(\overline{\mathbf{X}}) - \hat{\boldsymbol{\mu}}_l\|_2 + \|\boldsymbol{\sigma}_l(\overline{\mathbf{X}}) - \hat{\boldsymbol{\sigma}}_l\|_2 \right),
\end{equation}
where $\overline{\mathbf{X}} = \frac{\mathbf{X} - \boldsymbol{\mu}_{\text{D}}} {\boldsymbol{\sigma}_{\text{D}}}$ is the normalized input to the network. To avoid high-frequency noise, the pixel variation can also be optimized, promoting smooth areas in the data samples
\begin{multline}
\mathcal{L}_\text{TV}(\overline{\mathbf{X}}) = \sum_{b=0}^{B-1} \sum_{c=0}^{C-1} \sum_{h=0}^{H-2} \sum_{w=0}^{W-2} (X_{b,c,h+1,w} - X_{b,c,h,w})^2 + \\(X_{b,c,h,w+1} - X_{b,c,h,w})^2.
\end{multline}
And finally, considering the cross entropy (CE) loss with the random label vector $\mathbf{y}$, the full loss function becomes 
\begin{multline}
\mathcal{L}_{\text{total}}(\mathbf{X},\mathbf{y}) =\lambda_{\text{D}} \mathcal{L}_{\text{D}}(\mathbf{X}) + \lambda_{\text{BN}} \mathcal{L}_{\text{BN}}(\overline{\mathbf{X}}) + \\ \lambda_{\text{CE}}  \mathcal{L}_{\text{CE}}(\overline{\mathbf{X}},\mathbf{y}) + \lambda_{\text{TV}} \mathcal{L}_{\text{TV}}(\overline{\mathbf{X}}) ,
\end{multline}
where each $\lambda_i$ represents a hyperparameter regulating the strength of the losses. Each extracted batch, alongside their corresponding labels, can be used to BN-tune the attacked model without any prior main task data. An example of these images can be seen in Figure~\ref{fig:image_examples}.

\section{Implementation and Experimental Results}\label{sec:experimental}

\subsection{Reference works}

The efficacy of the proposed attack was validated experimentally against two different BN watermarking schemes: FedIPR~\cite{Bowen_Li22}, which embeds global watermarks into the BN parameters of Passport layers; and FedTracker~\cite{Shao23}, which uses the BN parameters of vanilla BN layers to embed unique watermarks into different model copies, identifying each data-owner in the system. The works in~\cite{Chen25,Yang26} also follow a strategy comparable to~\cite{Shao23}, and differ either in the encoding and embedding of the watermark~\cite{Chen25}, or the information embedded~\cite{Yang26}. As there are no public code repositories for these works, they were not included in this study.\footnote{We contacted the authors of these works to request access to their implementations, but did not receive a response.} Finally, the works in~\cite{WYang23,Chao24,Karim26} directly borrow their BN watermarking embedding from~\cite{Bowen_Li22}. Both evaluated methods were used to train multiple models with 20 data-owners, keeping the configuration close to the original work whenever possible:
\begin{itemize}
    \item FedIPR~\cite{Bowen_Li22}: AlexNet~\cite{AlexNet} on CIFAR-10~\cite{CIFAR10} and ResNet18~\cite{ResNet} on CIFAR-100~\cite{CIFAR10} for the IID experiments, and AlexNet on CIFAR-10 with $\beta=0.1$ for the non IID ones.
    \item FedTracker~\cite{Shao23}: VGG16~\cite{VGG} on CIFAR-10 and ResNet18 on CIFAR-100 for the IID experiments, and VGG16 on CIFAR-10 with $\xi=0.5$ for the non IID. For ResNet18, the watermark was hidden into the last two BN layers.\footnote{The explicit configuration for ResNet18 was not included in the original paper, but authors indicate that the watermark should be embedded in the deeper BN layers and not exceed a certain number of BN parameters to avoid overfitting to multiple data-owner's watermark.}
\end{itemize}

Both reference works originally include black-box watermarks along with the BN watermark under consideration. Because of the implementation of these watermarks, for the experiments with CIFAR-100 they were either limited to 10 classes in the case of FedIPR, or suppressed in the case of FedTracker, given that in the latter the BN layers are frozen during black-box embedding.

\subsection{Experimental Settings of the Attacks}

Unless stated otherwise, the proposed attack uses the linear layer-wise re-initialization and data-free BN-tuning described in Section~\ref{sec:proposed}. For the latter, the model trains on 100 batches of 16 samples, with the same configuration as the local training. Each batch is generated setting $\lambda_\text{D}=10$, $\lambda_\text{BN}=1$, $\lambda_\text{CE}=5$ and $\lambda_\text{TV}=10^{-4}$, though 2000 iterations. The initial 85\% of iterations include data augmentation transformations over the normalized $\overline{\mathbf{X}}$ (random crops with 4-pixel padding, random horizontal flips), and the learning rate is adjusted from $0.02$ to a minimum of $10^{-4}$ through a Cosine Learning Rate Scheduler. This configuration, alongside the values of $\lambda_i$, was found experimentally through visual inspection of the reconstructed images. Regardless of the possibility of detection presented in Section~\ref{subsec:detectability}, the attack is always performed across all compatible layers in the model (all BN layers for FedTracker, all Password Layers for FedIPR). This avoids assuming that the attacker has prior knowledge of the watermark location, which could potentially be key-dependent, and also ensures that the removal process is not compromised by an incomplete selection of layers.

In experiments using main task data, fine-tuning and BN-tuning is performed for 50 epochs of a local data-owner's dataset, with the original training settings in each case. For naive BN-tuning, models trained with FedTracker use a learning rate 100 times higher than the original ($1.0$), while those trained with FedIPR use a learning rate 10 times higher ($0.1$), to perturb the BN parameters as much as possible without destroying the main-task performance.

All attacks were evaluated across 5 attempts by different data-owners, in terms of main task accuracy on the test dataset, and watermark bit accuracy. For FedIPR, where multiple watermarks coexist within the same global model, the reported bit accuracy was averaged across all watermarks; for FedTracker, it corresponds to the guilty participant's watermark.

\subsection{Experimental Results}

\subsubsection{Comparison to Naive Data-Driven Attacks}

\begin{table*}[h]

\caption{Performance comparison across different attacks, in terms of main task accuracy preservation and bit accuracy of the watermark. The best results after the attacks are highlighted in each case.}
\begin{tabular}{rr|cccc|cccc|}
\cline{3-10}
\multicolumn{1}{l}{}                              & \multicolumn{1}{l|}{} & \multicolumn{4}{c|}{Main Task Accuracy (\%)}                                                                            & \multicolumn{4}{c|}{Watermark Bit Accuracy (\%)}                                                                                  \\ \cline{3-10} 
\multicolumn{1}{l}{}                              & \multicolumn{1}{l|}{} & \multicolumn{1}{c|}{Before Attack} & \multicolumn{1}{c|}{Fine-tuning} & \multicolumn{1}{c|}{BN-tuning} & Ours           & \multicolumn{1}{c|}{Before Attack} & \multicolumn{1}{c|}{Fine-tuning} & \multicolumn{1}{c|}{BN-tuning} & Ours           \\ \hline
\multicolumn{1}{|r|}{\multirow{2}{*}{FedIPR~\cite{Bowen_Li22}}}     & AlexNet w/ CIFAR-10    & \multicolumn{1}{c|}{89.94}         & \multicolumn{1}{c|}{75.08}       & \multicolumn{1}{c|}{83.80}          & \textbf{88.17} & \multicolumn{1}{c|}{95.69}         & \multicolumn{1}{c|}{66.04}       & \multicolumn{1}{c|}{65.74}          &  \textbf{52.37}\\ \cline{2-10} 
\multicolumn{1}{|r|}{}                            & ResNet18 w/ CIFAR-100  & \multicolumn{1}{c|}{71.74}         & \multicolumn{1}{c|}{45.65}       & \multicolumn{1}{c|}{61.25}          & \textbf{70.03} & \multicolumn{1}{c|}{94.25}         & \multicolumn{1}{c|}{58.30}       & \multicolumn{1}{c|}{64.40}          & \textbf{50.40} \\ \hline
\multicolumn{1}{|r|}{\multirow{2}{*}{FedTracker~\cite{Shao23}}} & VGG16 w/ CIFAR-10      & \multicolumn{1}{c|}{89.59}         & \multicolumn{1}{c|}{79.68}       & \multicolumn{1}{c|}{86.48}     & \textbf{87.62} & \multicolumn{1}{c|}{96.56}         & \multicolumn{1}{c|}{95.31}       & \multicolumn{1}{c|}{76.41}     & \textbf{49.22} \\ \cline{2-10} 
\multicolumn{1}{|r|}{}                            & ResNet18 w/ CIFAR-100  & \multicolumn{1}{c|}{73.67}              & \multicolumn{1}{c|}{\textbf{73.61}}            & \multicolumn{1}{c|}{66.88}          &        72.46        & \multicolumn{1}{c|}{97.03}              & \multicolumn{1}{c|}{94.53}            & \multicolumn{1}{c|}{70.31}          &         \textbf{50.47}       \\ \hline
\end{tabular}\label{tab:general_results}
\end{table*}

Looking at how the proposed data-free removal compares to the naive attacks in Table~\ref{tab:general_results}, the former outperforms them in both preserving the main-task accuracy and removing the watermark across most experiments. It is particularly significant that the average watermark bit accuracy approaches random guessing, giving the adversary a dangerous advantage:\textbf{ a guarantee of success}, as no detector would be able to perform an accusation against them. As with any other attack, the removal cannot be verified without access to the secret key, which would normally be sufficient to deter a malicious participant if the probability of success were not too high~\cite{Skoric12}. However, knowing that the attack is capable of completely removing the watermark eliminates this psychological deterrent.

\subsubsection{Using Representative Main Task Data}

\begin{table}[h]
\caption{Effect of using representative main task data during BN-tuning in the proposed attack. The best results are highlighted in each case. }
\centering
\begin{tabular}{rr|cc|cc|}
\cline{3-6}
\multicolumn{1}{l}{}                                                       & \multicolumn{1}{l|}{} & \multicolumn{2}{c|}{Main Task Acc (\%)}     & \multicolumn{2}{c|}{Watermark Bit Acc (\%)}             \\ \cline{3-6} 
\multicolumn{1}{l}{}                                                       & \multicolumn{1}{l|}{} & \multicolumn{1}{c|}{Local Data}     & Data-free  & \multicolumn{1}{c|}{Local Data} & Data-free            \\ \hline
\multicolumn{1}{|r|}{\multirow{2}{*}{\cite{Bowen_Li22}}} & AlexNet    & \multicolumn{1}{c|}{\textbf{88.48}}               & 88.17 & \multicolumn{1}{c|}{59.10}           & \textbf{52.37} \\ \cline{2-6} 
\multicolumn{1}{|r|}{}                                                     & ResNet18  & \multicolumn{1}{c|}{69.59}               & \textbf{70.03} & \multicolumn{1}{c|}{56.35}           & \textbf{50.40} \\ \hline
\multicolumn{1}{|r|}{\multirow{2}{*}{\cite{Shao23}}}      & VGG16      & \multicolumn{1}{c|}{\textbf{89.46}} & 87.62 & \multicolumn{1}{c|}{52.97}      & \textbf{49.22} \\ \cline{2-6} 
\multicolumn{1}{|r|}{}                                                     & ResNet18  & \multicolumn{1}{c|}{\textbf{73.14}} & 72.46 & \multicolumn{1}{c|}{50.94}      & \textbf{50.47} \\ \hline
\end{tabular}\label{tab:representative_data}
\end{table}

While this work attempts to prove the potential of a data-free attack, an adversary with representative main task data could instead leverage their local dataset. The comparison to this data-driven alternative is presented in Table~\ref{tab:representative_data}. As expected, using representative data generally leads to better preservation of the main-task performance, but it also results in a slight recovery of the watermark. This could be due to the fact that the local datasets guide the model back toward the original local minimum. Considering this, and depending on the attacker's objective, the data-free attack may be a better option even with a greater loss of performance.

\subsubsection{Impact of non i.i.d. Settings}\label{subsec:exp_noniid} 
\begin{figure}[ht]
    \centering

    \subfloat[VGG16 with FedTracker trained on non IID CIFAR-10]{\includegraphics[width=\columnwidth]{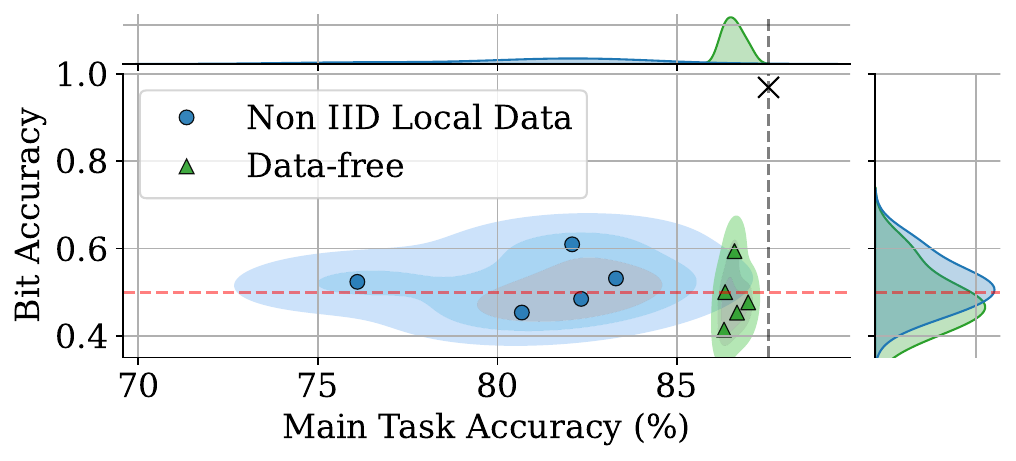}}\\
    \subfloat[AlexNet with FedIPR trained on non IID CIFAR-10]{\includegraphics[width=\columnwidth]{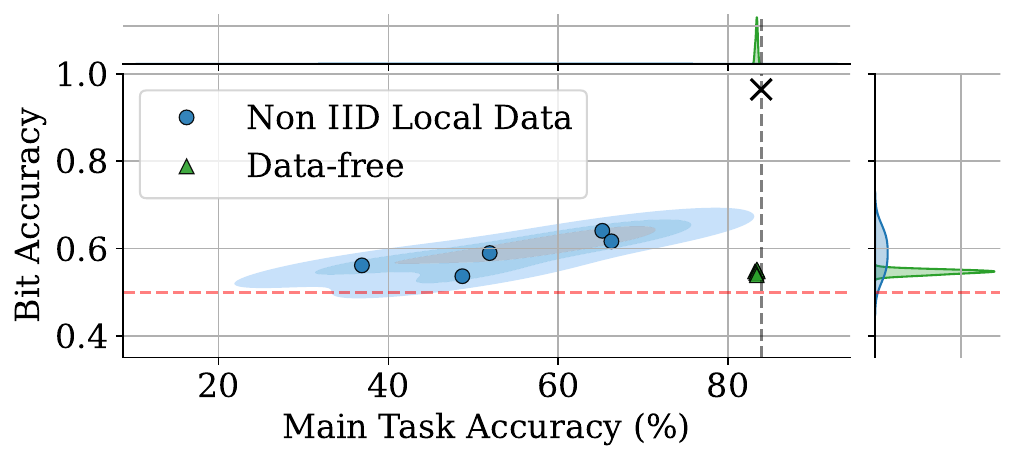}}\\
    \caption{Comparison of our data-free BN-tuning against non IID training data. Dashed lines represent the random guessing of the watermark (red) and original performance of the model before the attack (gray).}
    \label{fig:noniid}
\end{figure}

While it may be a common assumption that a participating FL data-owner can use their local main task data to tune the stolen model, real life FL applications are typically more complex, and using non IID data might be a self-defeating solution. Figure~\ref{fig:noniid} analyzes the most challenging non-IID settings considered in the original schemes, generated using a Dirichlet distribution with concentration parameter $\beta=0.1$ for FedIPR and $\xi=0.5$ for FedTracker. Under these conditions, the proposed data-free BN-tuning still performs adequately compared to the unpredictability of the local main task datasets.

\subsubsection{Impact of a Random Initialization}\label{subsec:exp_initialization} 
\begin{figure}[ht]
    \centering
    \includegraphics[width=\columnwidth]{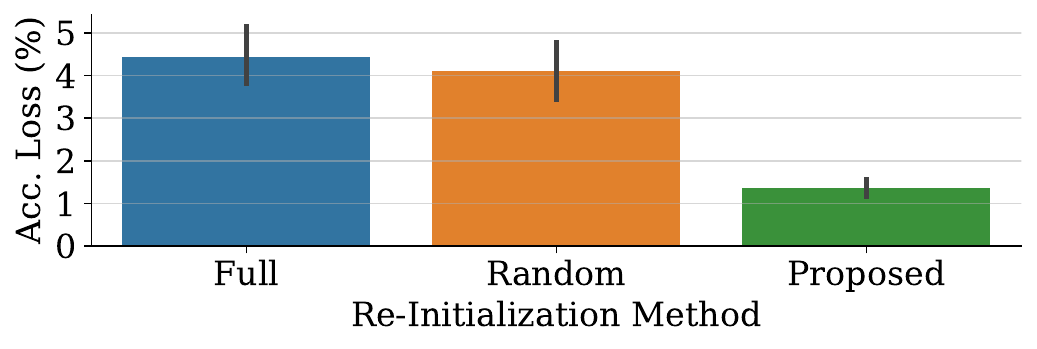}
    \caption{Comparison of re-initialization methods for the BN parameters in terms of accuracy loss after the data-free attack.}
    \label{fig:reinitialization}
\end{figure}

Although the BN-tuning step is meant to find a local minimum for the classification, the first initialization step based on batch statistics is still critical to minimize the impact to the model's performance. Figure~\ref{fig:reinitialization} compares the main task accuracy loss for all models trained in this work, after alternative re-initializations that do not involve batch statistics: Full re-initialization, restoring $\boldsymbol{\gamma}_l = \mathbf{1}_{C}$ and $\boldsymbol{\beta}_l = \mathbf{0}_{C}$; and random re-initialization, as $\boldsymbol{\gamma}_l \sim \mathcal{N}\left(\operatorname{mean}(\boldsymbol{\gamma}_l^\text{orig}), \, \operatorname{std}(\boldsymbol{\gamma}_l^\text{orig})^2\right)$, using the original BN parameters to generate vectors $\boldsymbol{\gamma}_l$ and $\boldsymbol{\beta}_l$ of the same characteristics. While a random re-initialization performs better than disregarding the trained parameters altogether, the proposed approach based on running batch statistics (Section~\ref{sec:proposed}) makes the best use of the information learned by the network and consistently achieves the smallest performance degradation.

\subsubsection{Reconstructed Main Task Samples} 
\begin{figure}[ht]
    \centering
    \subfloat[airplane]{\includegraphics[width=0.19\columnwidth]{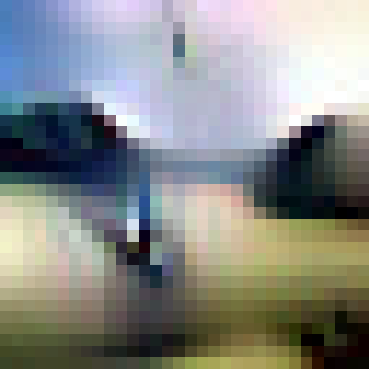}}\hfill
    \subfloat[automobile]{\includegraphics[width=0.19\columnwidth]{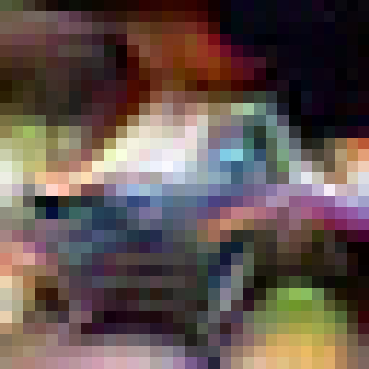}}\hfill\subfloat[bird]{\includegraphics[width=0.19\columnwidth]{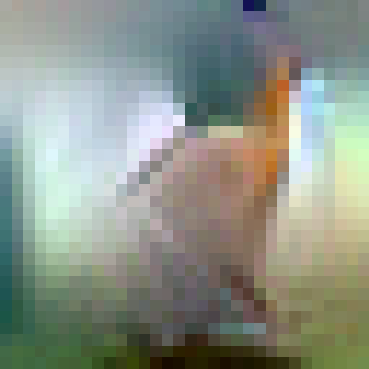}}\hfill\subfloat[cat]{\includegraphics[width=0.19\columnwidth]{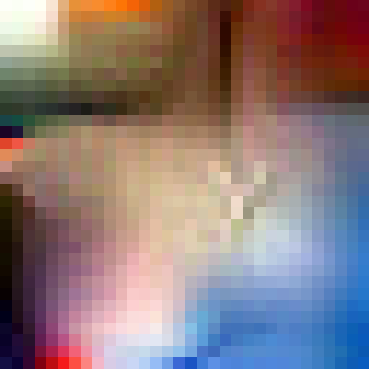}}\hfill\subfloat[deer]{\includegraphics[width=0.19\columnwidth]{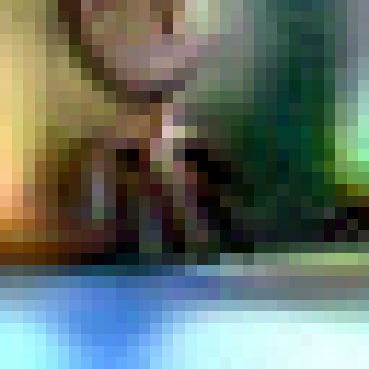}}\\
     \subfloat[dog]{\includegraphics[width=0.19\columnwidth]{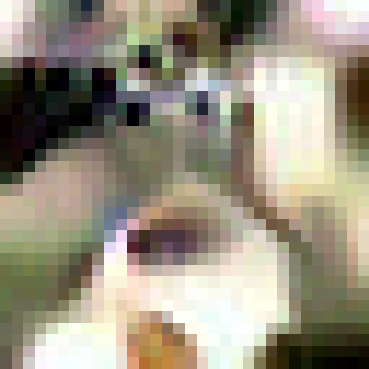}}\hfill
    \subfloat[frog]{\includegraphics[width=0.19\columnwidth]{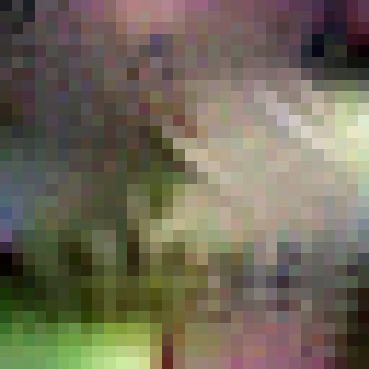}}\hfill\subfloat[horse]{\includegraphics[width=0.19\columnwidth]{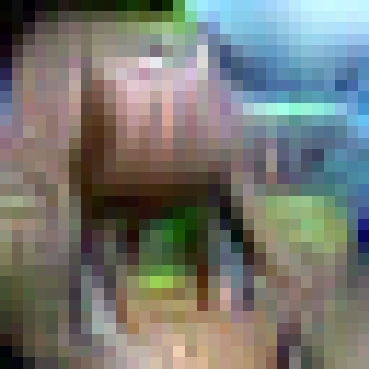}}\hfill\subfloat[ship]{\includegraphics[width=0.19\columnwidth]{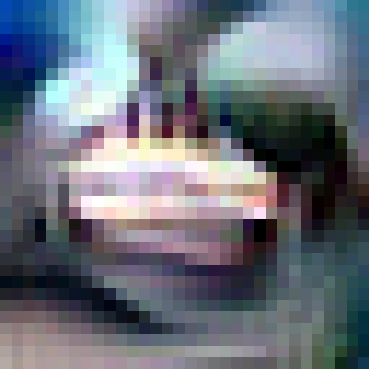}}\hfill\subfloat[truck]{\includegraphics[width=0.19\columnwidth]{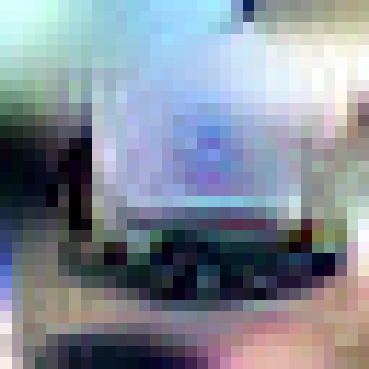}}\\
    \caption{Examples of reconstructed data for VGG16 trained on CIFAR-10 with FedTracker.}
    \label{fig:image_examples}
\end{figure}

An example of reconstructed samples for the proposed data-free BN-tuning can be seen in Figure~\ref{fig:image_examples}. Although these images are clearly of poor visual quality and would likely not be suitable to train a surrogate model from scratch, such fidelity is not required for BN-tuning. Since the attack only adjusts the linear BN parameters while leaving the rest of the network unchanged, it has a limited impact on the learned main-task function and is less likely to overfit to reconstruction artifacts. Nevertheless, more advanced data distillation techniques could further reduce the performance loss, potentially enabling a final fine-tuning step over the entire network.

\section{Conclusions}\label{sec:conclusions}

This paper analyzed the vulnerabilities of BN watermarks in FL, identifying previously overlooked weaknesses and demonstrating how they can be exploited through a practical data-free removal attack. Across two representative schemes, the proposed attack consistently removed the watermark while mostly preserving the accuracy of the stolen model.

Beyond its empirical performance, this attack has important implications for the security of current BN-based schemes, as the high success rate effectively eliminates the uncertainty of detection that would otherwise discourage model theft. More generally, these results reinforce the need to evaluate watermarking schemes under informed threat models, following Kerckhoffs's principle, rather than relying on security through obscurity or robustness against generic model modifications.

Ultimately, these findings suggest that watermarking schemes should avoid embedding into components that can be isolated and replaced. Instead, watermark information should be tightly entangled with the main task function, so that its removal necessarily degrades model performance. Two alternative strategies to achieve this entanglement appear promising: One is to embed the watermark directly into important parameters~\cite{Namba19,RodriguezLois25}, accepting a potentially greater impact on the main task; the other is to avoid perturbing those key parameters while concealing the effective embedding locations, through techniques such as Wet Paper Coding~\cite{Wang22} and unlearning analysis~\cite{Pan25}.

\bibliographystyle{ieeetr}
\bibliography{bibliography}
\end{document}